# A Diffractive Neural Network with Weight-Noise-Injection Training

**Jiashuo Shi[1]   Xinyu Zhang[1,2]**

[1] National Key Laboratory of Science and Technology on Multispectral Information Processing, Huazhong University of Science and Technology, Wuhan 430074, China

[2]Corresponding author

E-mail:shijiashuo97@163.com

## Abstract

We propose a diffractive neural network with strong robustness based on Weight-Noise-Injection training, which achieves accurate and fast optical-based classification while diffraction layers have a certain amount of surface shape error. To the best of our knowledge, it is the first time that using injection weight noise during training to reduce the impact of external interference on deep learning inference results. In the proposed method, the diffractive neural network learns the mapping between the input image and the label in Weight-Noise-Injection mode, making the network's weight insensitive to modest changes, which improve the network's noise resistance at a lower cost. By comparing the accuracy of the network under different noise, it is verified that the proposed network (SRNN) still maintains a higher accuracy under serious noise.

## 1 Introduction

Discriminant learning, especially artificial neural networks, has drastically impacted the area of image recognition, image classification, and image super-resolution in recent years[12-16]. As the number of parameters and connections in neural networks has grown dramatically, the computing capability of traditional central processing units(CPUs) can't meet the demand. Although much other computing hardware, including IBM's TrueNorth chip, Google's tensor processing units (TPUs) and graphical processing units(GPUs), has been developed to improve the processing speed, their computing speeds are limited by electron mobility and operating frequency.

Since light has advantages of broad bandwidth, high transmission speed, and low crosstalk, it can overcome the bottleneck of the speed for traditional electric computing hardware. Based on these strengths, scientists have mainly proposed two optical neural network structures in recent years. The first relies on silicon photonic, which realizes matrix multiplication by an optical on-chip platform mainly consisted of several Mech-Zender interferometers(MZI)[17,18]. Another neural network is physically formed by multiple layers of diffractive surfaces that work in collaboration to perform an arbitrary function optically[2]. These all-optical neural networks make linear transformations can be implemented at the speed of light and detected at rates exceeding 100 GHz[1].

For the diffraction optical neural network based on the Huygens-Fresnel principle, many versions have appeared. After scientists first proposed a passive diffraction neural network based on 3D printing to successfully classify MNIST and Fashion-MNIST in terahertz band [2-4], JulieChang et al. successfully realized the optical convolution operation of single layer by adding an optimized convolution kernel to the phase modulation plane of the 4-f system [5], Ying Zuo et al. used Fourier lenses and spatial light modulators instead of a fixed printing structure to achieve an optical network with intensity as transfer parameter, and introduced optical nonlinear activation functions through laser-cooled atoms with electro-magnetically induced transparency [6].

The above-mentioned various diffractive neural networks rely on the computer to complete the backpropagation for optimizing the learnable parameter of networks and then realize the forward propagation of ultra-low energy consumption through the single-layer or multi-layer diffractive phase plate to realize classification or imaging. Since there is no consideration of errors, the parameters obtained by computer are ideal results, which are susceptible to error interference under all optical conditions. For all-optical diffraction neural network(DNN) of the 3D printing structure, the error sources mainly originate in the following aspects: (1)The lack of precision and accuracy of the 3D-printer. (2)The edges of the phase plate are not aligned. (3) The phase mask is exceedingly thin, which causes a slight deformation in the actual installation. (4)The abrasion of the phase plate. (5)The layer spacing error. (6)The slight change in the frequency of the terahertz light source. To mitigate the impact of these errors, one can improve hardware accuracy and maintain experimental devices, however, enhancing the implementation strategy of the simulation process is a more convenient and cheap way obviously. This paper mainly reduces the influence of (1)(4)(5)(6) error from the perspective of backpropagation. After experimental verification, a strong robust diffractive neural network is realized, using the method of Weight-Noise-Injection training, which pushes the model into the vicinity of the minimum value surrounded by the gentle area. It should be noted that Weight-Noise-Injection has been applied to neural networks such as RBF network, MLPs and recurrent neural networks to improve the convergence ability, generalization and fault tolerance of neural networks [7-9] while this study aims to improve the resistance of DNN to multiple types of actual hardware errors, rather than artificial weight errors[9], i.e., let the DNN maintain high-precision prediction with hardware errors. In theory, the DNN can be trained according to the type of error in the actual environment to achieve high-precision prediction under any type of error. The contributions of our work are as follows:

We propose a strong robustness diffraction neural network(SRNN) model, which mitigates the effect of phase errors on DNN output in a non-hardware way at a lower cost. By adding different standard deviations of Gaussian noise to the weights during training, the error distribution of the actual phase mask was simulated to make the optimized DNN become more resistant to phase errors. In other words, the Weight-Noise-Injection training forces the optimal weights to find a region of the minimum that is relatively insensitive to errors instead of only performing SGD to a minimum value of the loss function.

Then we evaluated the effect of the (1)(4)(5)(6) on traditional diffraction neural network(DNN) and SRNN, it is found that our proposed neural network model has a staggering advantage in dealing with these errors.

## 2 The proposed method

For the purely phase-modulated diffraction neural network, the propagation of optical waves

could be described as follow:

$$m_i^l = 1 \cdot e^{j\varphi_2} \tag{1}$$

$$w_i^l = \frac{z^l - z^{l-1}}{r^2}(\frac{1}{2\pi r} + \frac{1}{j\lambda})\exp(\frac{j2\pi r}{\lambda}) \tag{2}$$

$$output_i^l = w_i^l \times \sum\nolimits_k output_k^{l-1} \times m_i^l = w_i^l \times |A|e^{j\varphi_1} \times e^{j\varphi_2} = |A_w|e^{j\Delta\varphi} \tag{3}$$

Where $output_i^l$ is the output wave to $i$-th neuron of layer $l$, $m_i^l$ is the modulation of individual neurons, $w_i^l$ is the Rayleigh-Sommerfeld diffraction between neuron in layer $l$ and neuron in layer $l-1$, $r$ denotes distance between two neurons.

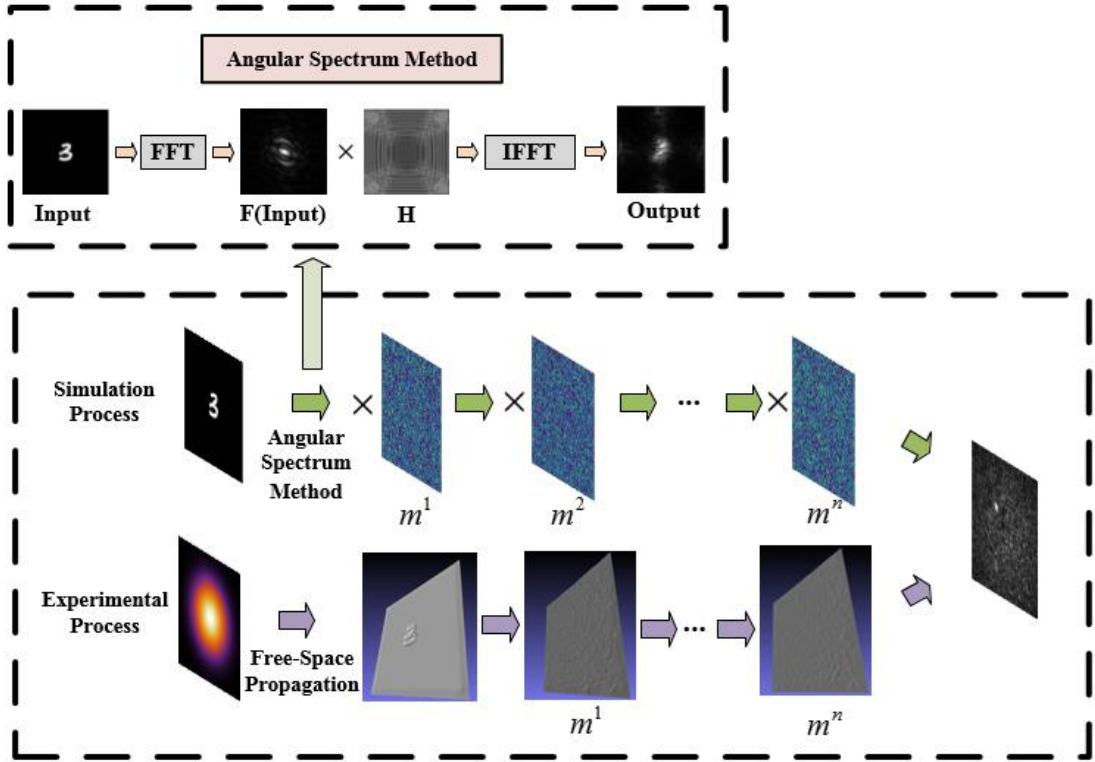

Figure 1 the DNN simulation and experimental process

Figure 1 is the forward propagation process of DNN in the simulation and experimental environment, where the simulation process uses the Fourier Transform to calculate the near-real diffraction effect based on the angular spectrum method. $m^l$ represents the $l$ th diffraction layer, H is the constant diffraction transfer function representing the Rayleigh-Sommerfeld diffraction formula. And each green arrow indicates the calculation of the angular spectrum method to obtain the complex amplitude distribution on the next diffraction plane. In fact, the actual diffraction neural network uses a Gaussian light source to project on different data amplitude modulation masks for obtaining the complex amplitude distribution of the data. The input light diffracts in free space and is coded in each diffraction layer to achieve classification or imaging, etc. In addition, the noteworthy difference between the two processes is that the computer runs a noise-free light propagation process.

2.1 Selection of loss function

This study uses cross-entropy as the loss function of the diffractive neural network. Unlike [3], we use the following loss function:

$$L(p,q) = -\sum_{x}(p(x)\log q(x) + (1-p(x))\log(1-q(x))) \quad (4)$$

Where $p$ is the expected probability distribution, $q$ is the probability distribution of network output. This loss function hopes that the probability distribution of the options other than the correct option will be more uniform and tend to a smaller value, which further reduces the probability of a higher voting result of a wrong option. In other words, the ideal situation should be to maximize ground truth while also making the proportion of other options more average, instead of focusing on a certain error option.

2.2 The injection weight noise strategy for DNN

Consider a diffractive neural network is trained using injection weight noise strategy, the network update equation will be given by

$$\varphi(s+1) = \varphi(s) - \mu \frac{\partial(y_t - output(x_t, \varphi_n(s)))}{\partial \varphi} \quad (5)$$

Where $\varphi$ (the phase value of diffraction mask) is the learnable parameter, $\varphi_n$ is the noise phase, which is expressed as $\varphi_n(s) = \varphi(s) + noise$, $s$ denotes $s$-th iteration of neural network, $(x_t, y_t)$ is the training set, $\mu$ reprensents learning rate.

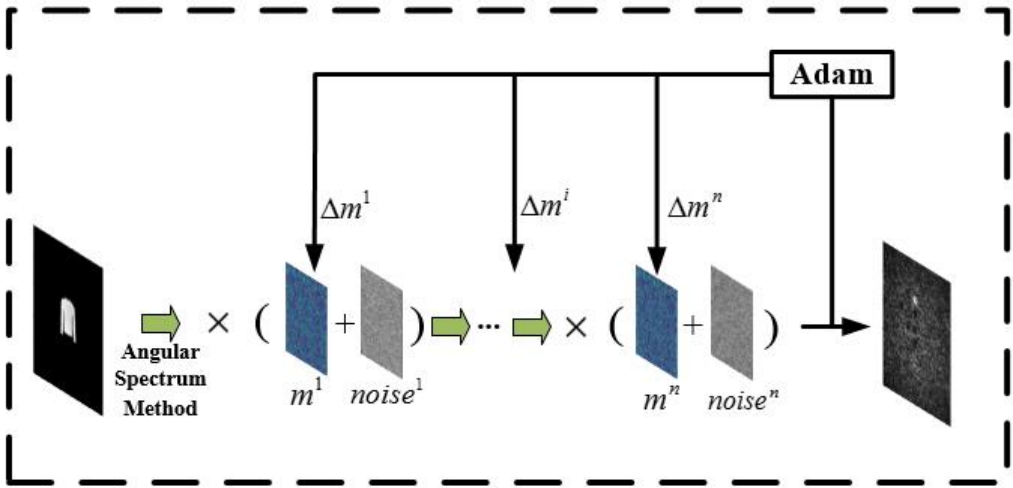

Figure 2 The SRNN back-propagation model

Figure 2 shows the SRNN back-propagation model. Compared with the traditional DNN, the proposed model needs to add a certain proportion of random noise matrix to each diffraction layer based on Gaussian distribution when using the data set for network training. It is known that most of the noise can be approximated as Gaussian noise, hence we use Gaussian noise instead of other noise injects to network weight. Of course, other errors can be injected into the weight, but they are not necessarily as universal as Gaussian noise. For each iteration, these noise matrices are randomly and independently generated and added to the network weights, which add a certain component of randomness to the phase modulation process of each layer, causing the output deviation of diffraction neural network. We have verified by experiments that this training method pushes the model into the vicinity of the minimum value surrounded by the gentle area, which means the weight of the network will be insensitive to modest errors.

## 3 Results and discussion

Figure 3 is the process of training DNN and SRNN separately using the MNIST, the number in the label of SRNN represents the standard deviation of the random noise injected to the model during the backpropagation. It should be noted that although the standard deviations are distinct, the mean of the random matrices is zero. It can be seen that the larger standard deviation of the added noise, the slower network training, and the lower accuracy. This is because the larger noise will add severe randomness to the network output at the beginning until the weight of the neural network entering a flat area that is weakly affected by small changes, in a sense, the neural network can overcome the impact of the injected noise. To ensure the rigor of subsequent experiments, we train each neural network 20 epochs using the same data set. The number of neurons in each layer(five layers) is 200, with a spacing of 3cm, using a 400GHz light source.

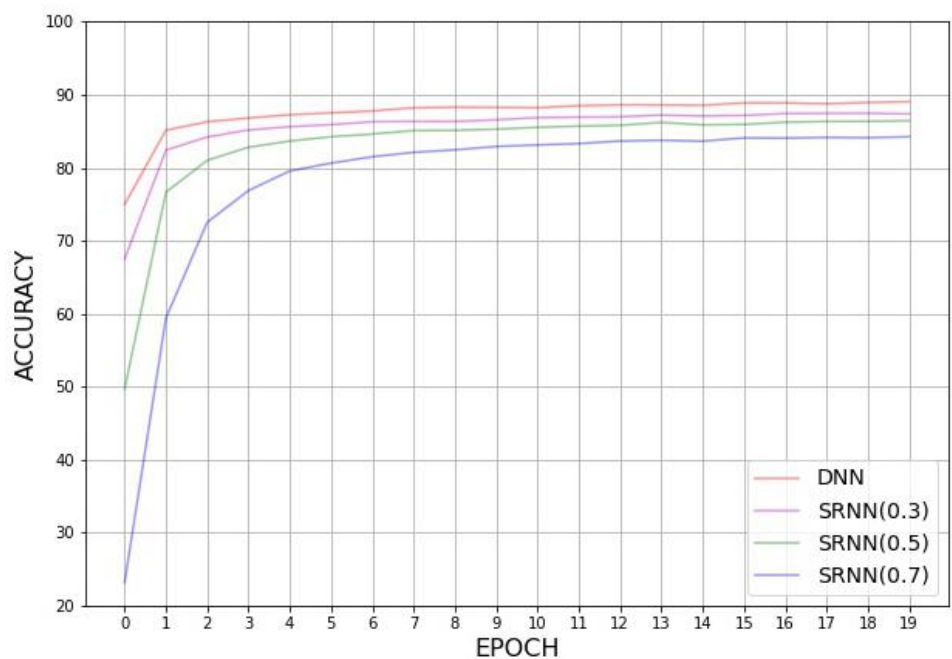

Figure 3 DNN and SRNN on MNIST

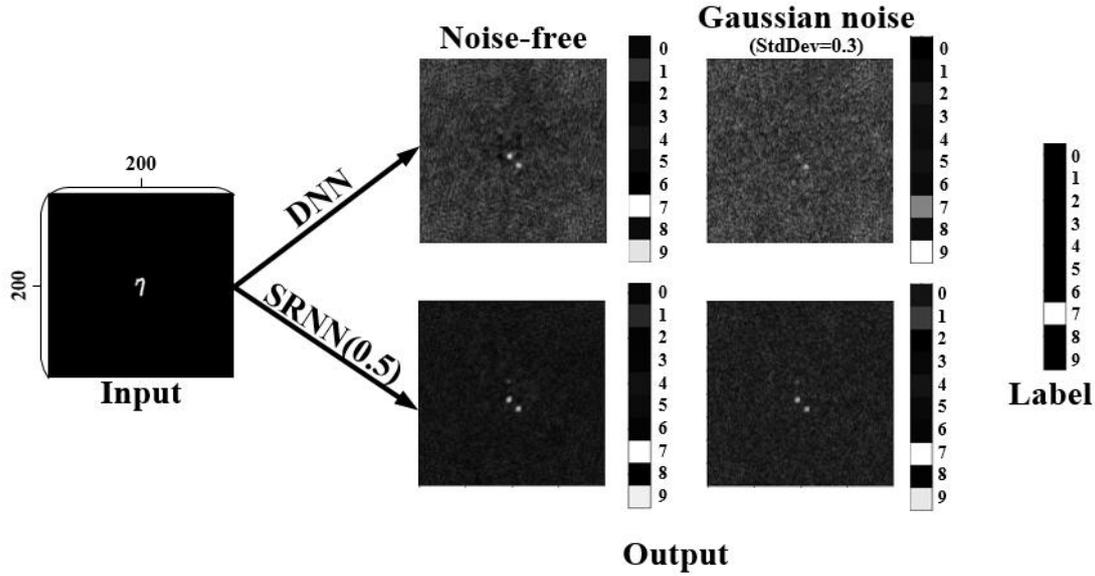

Figure 4  The classification of number 7 under noise-free and gaussian noise separately

As shown in Figure 4, we compare the predictions of the two trained networks under noise-free and Gaussian noise(Mean=0,StdDev=0.3). Obviously, the model we proposed has better robustness to random noise. In Figure 5, histograms show the effect of different errors on trained DNN and SRNN on the test set. In (a), we inject the Gaussian matrix generated by the same random seed to each network weights and calculate the classification accuracy, eliminating the doubt that Figure 4 is a special case. For each standard deviation of the Gaussian matrix, we select 12 groups, the absolute accuracy is the average of 12 results. It can be observed that SRNN is injected with a Gaussian random matrix during the training process, which causes it to be less affected by noise. On the contrary, the neural network classification is on the brink of random classification, when the standard deviation of the injected noise of the DNN reaches 1. In (b), we further analyze the effect of 3D printing Z-axis precision on the neural networks. It is known that the hardware implementation of the optical diffraction neural network relies on 3D printing the computer-optimized layer, which means phase modulation capability of each layer is severely affected by 3D printer Z-axis precision. In the experiment, we use the Z-axis precision of 0.1mm, 0.2mm, 0.4mm, 0.5mm to test the classification effect of these networks. It can be obviously seen that SRNN (0.3) has a lower classification accuracy than DNN at 0.1 mm, however, it shows impressive classification accuracy at a lower precision. Actually the SRNN training is injected with random Gaussian noise, but also has notable resistance to 3D printing errors, which don't belong to the Gaussian distribution. The best proof is the accuracy of DNN to 0.5mm is decreased by 35.4% compared to 0.1mm, while SRNN (0.3) is only decreased by 8.2%. The experimental results show that it is possible to use a low-precision 3D printer to make a lower-cost optical diffractive neural network, which has impressive robustness to random errors. (c)(d) further illustrate the injection Gaussian noise training is a generalization method, which shows the impact of changes in light source frequency and layer spacing on the experiment. It should be emphasized that these networks are trained under the conditions of a 3cm layer spacing and a 400GHz light source frequency. In (d), the random(2.9,3.1) means we use five randomly generated values in this range to set the layer spacing.

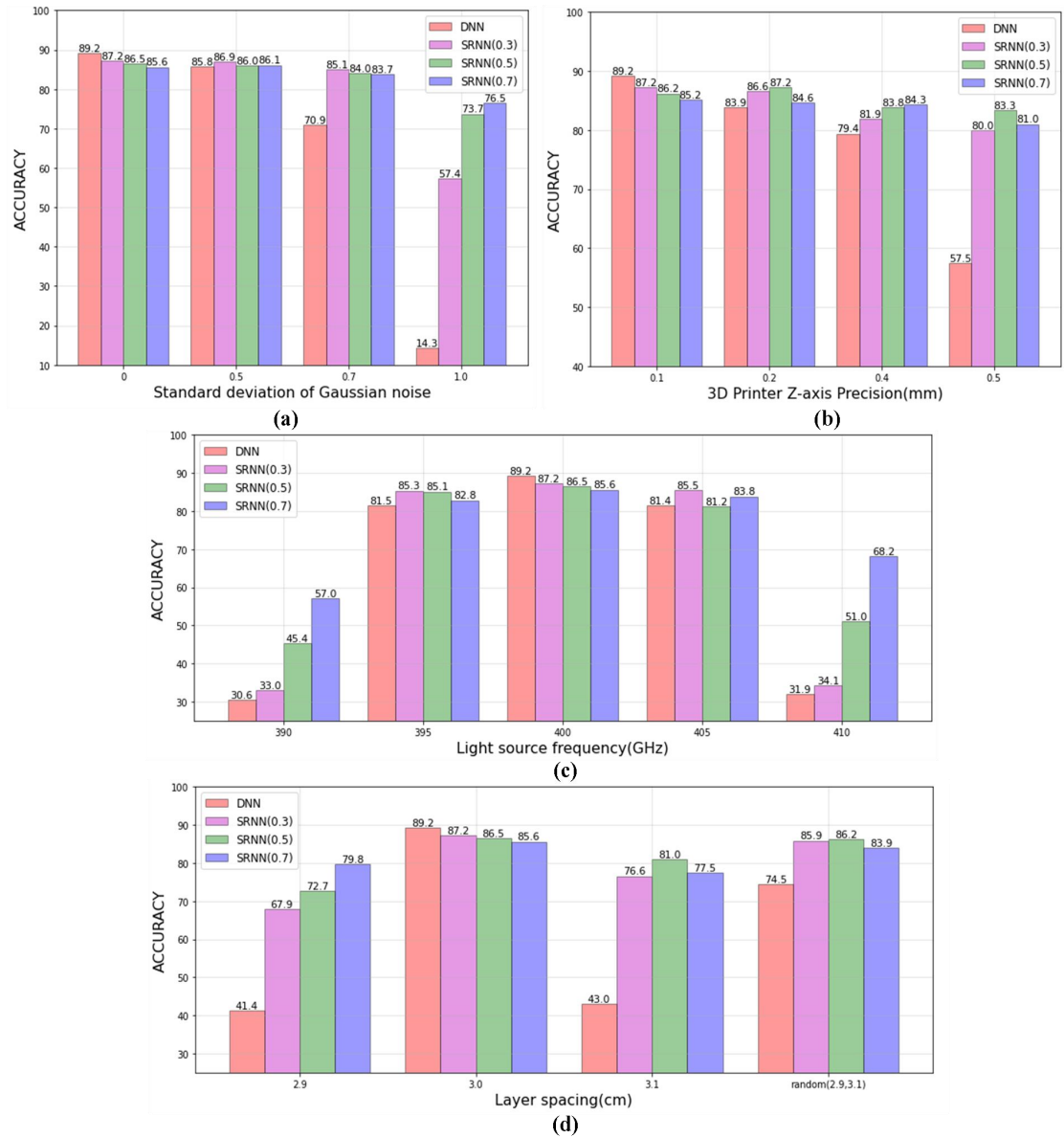

Figure 5　The effect of different errors on DNN and SRNN

The weight noise injection strategy is similar to the multiplying noise strategy of Dropout, which is to inject a fixed-scale additive noise into the network, they can all be regarded as a form of damage to the network structure, instead of the original input value. Generally, the application of weight noise injection on common networks will ill-conditioned prompt the network weight large to offset the fixed-scale additive noise(Relatively reduce noise). Nevertheless as shown in Figure 6(Where Height(m) in (c) has a fixed linear relationship with the network weights), for the special network structure without a linear rectification unit, such as DNN, weight noise injection does not bring about ill-conditioning, which means the weight noise injection is very effective for improving the robustness of this unconventional network structure by forces the optimal weights to find a region of the minimum that is relatively insensitive to errors.

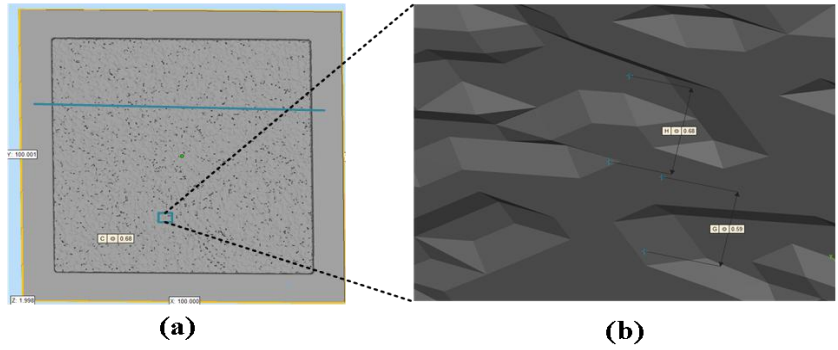
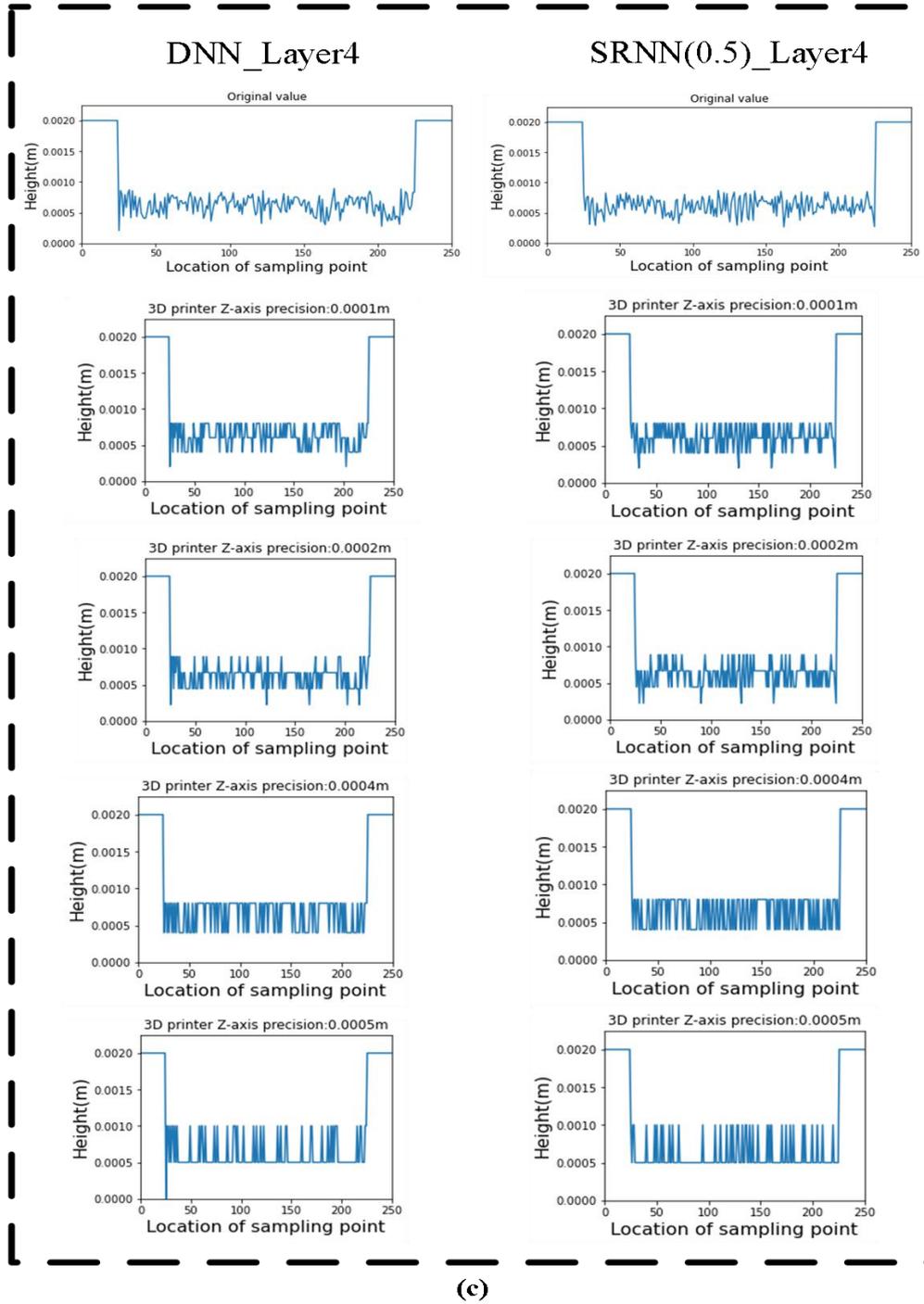

Figure 6 The network weights under different 3D printer precision

# 4 Conclusion

We propose a strong robustness diffraction neural network(SRNN) model, which mitigates the effect of phase errors on DNN output in a non-hardware way at a lower cost. By adding different standard deviations of noise to the weights during training, the error distribution of the actual phase mask was simulated to make the optimized DNN become more resistant to phase errors. By judging the effects of various errors on the two network models, we verify that our proposed model has more satisfactory robustness.